\documentclass[twocolumn,nonbibnotes,nofootinbib,notitlepage,longbibliography]{revtex4-2}
\usepackage{amsmath}
\usepackage{amssymb}
\usepackage{bm}
\usepackage{epsfig}
\usepackage{graphicx}
\usepackage{color}
\usepackage[dvipsnames]{xcolor}
\usepackage[colorlinks=true,citecolor=LimeGreen,linkcolor=blue, urlcolor = cyan]{hyperref}
\usepackage[english]{babel}

\begin{document}

\title{Ultrafast exciton transport in van der Waals heterostructures} 

\author{M.M. Glazov} 
\email{glazov@coherent.ioffe.ru} 
\affiliation{Ioffe Institute, 194021, St. Petersburg, Russia} 

\author{R.A. Suris} 
\affiliation{Ioffe Institute, 194021, St. Petersburg, Russia} 

\begin{abstract}
Excitons in van der Waals heterostructures based on atomically thin transition metal dichalcogenides are considered as potential candidates for the formation of a superfluid state in two-dimensional systems. A number of studies reported observations of ultrafast nondiffusive propagation of excitons in van der Waals heterostructures, which was considered by their authors as possible evidence of collective effects in excitonic systems. In this paper, after a brief analysis of exciton propagation regimes in two-dimensional semiconductors, an alternative model of ultrafast exciton transport is proposed, based on the formation of waveguide modes in van der Waals heterostructures and the radiation transfer by these modes.

\emph{Submitted to the special issue of Journal of Experimental and Theoretical Physics devoted to 130th anniversary of P.L. Kapitza.}
\end{abstract}

\maketitle

\section{Introduction}

The discovery~\cite{Kapitza:1938aa} in 1937 by Pyotr Leonidovich Kapitza  of  liquid helium superfluidity (see also~\cite{ALLEN:1938aa}) has made a strong impact on the development of condensed matter physics. Together with the phenomenological theory of this effect proposed by L.D.~Landau~\cite{landau:super}, as well as with an exactly solvable model of a weakly nonideal Bose gas, proposed by N.N.~Bogolyubov~\cite{bogolubov47:eng}, and microscopic theories of R.~Feynman~\cite{PhysRev.91.1291}, O.~Penrose and L.~Onsager~\cite{Penrose:1956zr}, which established, in particular, the connection between the superfluidity and Bose-Einstein statistics of $^4$He atoms (see, e.g.,~\cite{pitaevskii2003bose}), Kapitza's remarkable experiment has led to the formation of a new field of research and to an active search for other condensed matter systems where the Bose-Einstein condensation and superfluidity are possible.

In semiconductors and insulators, the Coulomb interaction between electrons and holes leads to formation of quasiparticles, excitons, which have integer spin and, accordingly, obey the Bose-Einstein statistics~\cite{PhysRev.37.17,PhysRev.52.191,Mott384}. After the experimental discovery of large-radius excitons in cuprous oxide~\cite{gross:exciton:eng} a number of authors drew attention to the possibility of condensation and therefore superfluidity of excitons in semiconductors~\cite{Moskalenko62:eng,PhysRev.126.1691,keldysh68a,suris_superfluid,1968JETP...26..104A}. In ideal two- or quasi-two-dimensional systems, where the free motion of particles is possible only in the plane, and motion in the transverse direction is quantized, Bose condensation in the strict sense of the term is impossible. However, at sufficiently low temperatures, a two-dimensional gas of repelling bosons experiences a Berezinskii-Kosterlitz-Thouless transition to a superfluid state~\cite{1971JETP...32..493B,1973JPhC....6.1181K}. Accordingly, a superfluid state is also possible for two-dimensional excitons~\cite{1975JETPL..22..274L}. Theoretically, non-diffusive transport of excitons in a thin semiconductor film was considered in Ref.~\cite{suris_superfluid} where by analogy with the a superfluid liquid flow in capillaries, the speed of a ring vortex with a characteristic diameter equal to the thickness of the film was used as the limiting speed. Experiments aimed at search and study of collective excitonic effects in quantum well structures, including those placed in microcavities, have been actively conducted for the last 30 years~\cite{PhysRevLett.73.304,0953-8984-16-50-R02, Gorbunov:2006aa, LeSiDang06,Christopoulos:2007kx,Lagoudakis:2008bh,Amo2009,Belykh:2012zr,Stern55,Shilo:2013aa,PhysRevLett.120.047402}, for reviews see~\cite{RevModPhys.85.299,microcavities,sanvitto_timofeev, GlazovSuris_2021}.

A rise of a novel material platform, van der Waals heterostructures based on transition metal dichalcogenides (TMDCs)~\cite{Geim:2013aa,Kolobov2016book}, whose optical spectra are controlled by Wannier-Mott excitons with a large ($100\ldots 500$~meV) binding energy, short ($\sim 1$~ps) radiative recombination time, effective coupling with light~\cite{Schneider:2018aa} and unusual fine structure of the energy spectrum and dynamics~\cite{Splendiani:2010a,Mak:2010bh,Selig:2016aa,RevModPhys.90.021001,Durnev_2018,Glazov_2021,Semina_2022}, has, naturally, aroused the interest of researchers in excitonic collective effects and their superfluidity in such systems~\cite{Fogler2014,PhysRevB.93.245410,PhysRevB.101.220504}. Despite significant efforts, to date, as far as we know, no unambiguous experimental evidence has been obtained for exciton superfluidity in TMDC-based heterostructures. The vast majority of experiments show the diffusive propagation of excitons in TMDC monolayers and van der Waals heterostructures based on these materials, see reviews~\cite{Chernikov:2023aa,Malic:2023aa}.

Nonetheless, in a number of works, an effective ``spreading'' of excitons in the plane of the structure with TMDC mono- and bilayers was observed~\cite{PhysRevB.104.165302,https://doi.org/10.48550/arxiv.2204.09760,fowlergerace2023transport,PhysRevB.107.045420}. In this regard the experiment of the Singapore group~\cite{Aguila:2023aa} is very remarkable. In this work the authors discovered that in a MoS$_2$-based channel the excitons propagate at a speed of $\approx 6~\%$ of the speed of light in vacuum $c=299792458$~m/s, and their distribution is well described by the model of an inviscid fluid. Such a scenario is, in principle, possible, but the propagation speed of excitons is too high: It exceeds all characteristic speeds that can be realized in two-dimensional semiconductors (including for superfluid excitons)~\cite{Glazov:2023aa}.

In this paper, after discussing the known in the literature scenarios for the exciton propagation in two-dimensional crystals (Sec.~\ref{sec:regimes}), we propose a model (Sec.~\ref{sec:model}),  which allows, at least at a qualitative level, to describe the ultrafast propagation of excitons in TMDC-based van der Waals heterostructures containing hexagonal boron nitride layers. In such structures, as estimates show, waveguide modes~\cite{Tamir:1988,Ebeling1993,39094} can be formed, propagating at high speed, only several times lower than the speed of light in vacuum. These modes can be excited, for example, by the recombination of photogenerated excitons and, propagating in the plane of the structure, can cause excitonic luminescence far from the excitation point. Thus, in principle, the propagation of waveguide modes may be responsible for the observed ultrafast exciton transport.

Note that the possibility of such ``secondary'' luminescence was mentionned earlier in connection with specifics of experimental data on exciton transport in now classical structures with GaN quantum wells~\cite{PhysRevApplied.6.014011}. The acceleration of the propagation of excitons in a TMDC monolayer in the strong coupling regime with the photonic modes of the substrate was discussed, for example, in Ref.~\cite{Guo:2022aa}, see also Ref.~\cite{PhysRevB.108.104202}, where the regimes of excitons and exciton-polaritons propagation in microcavities were analyzed in detail. The scenario we propose, however, refers to the weak coupling regime between an exciton in a two-dimensional semiconductor and a photon in a waveguide mode. We emphasize that even in this regime, where polaritons are not formed, multiple reflections of light from the interfaces of van der Waals heterostructures significantly affect the spectrum of excitons and their lifetime due to the Purcell and Lamb effects~\cite{PhysRevLett.120.037401,Horng:19,PhysRevLett.123.067401,ren2023control}. The formation of waveguide modes is also possible in other structures based on TMDCs, for example, in nanotubes~\cite{Eliseyev:2023aa}.

\begin{figure}[b]
 \includegraphics[width=\linewidth]{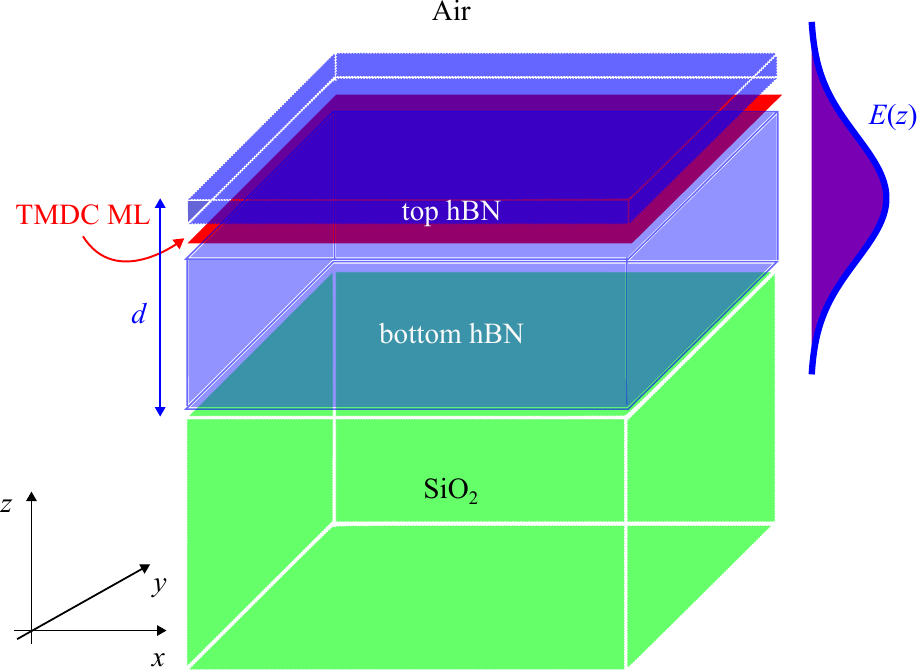}
  \caption{
Schematic illustration of van der Waals heterostructure with atom thin crystal (TMDC ML) encapsulated in hexagonal boron nitride (hBN) and deposited on the SiO$_2$ substrate (in many experiments a SiO$_2$ layer on silicon is used). A spatial profile of a fundamental waveguide mode in the structure is shown to the right.}\label{fig:Scheme}
\end{figure}

\section{Exciton propagation regimes}\label{sec:regimes}

A schematics of the van der Waals heterostructure based on TMDC studied in this work is shown in Fig.~\ref{fig:Scheme}. For example, a structure with a TMDC monolayer (TMDC ML) is shown, but all the analysis below applies to the structures with TMDC hetero- and homobilayers and, in principle, to structures with a larger number of atomically thin layers. Encapsulation of TMDC monolayers in hexagonal boron nitride (hBN) leads to a significant narrowing of exciton lines in all widespread materials: MoS$_2$, MoSe$_2$, WS$_2$ and WSe$_2$, and we discuss hBN encapsulated structures from the theory standpoint hereafter. Even at cryogenic temperatures the line broadening is mainly homogeneous~\cite{PhysRevX.7.021026}. For the ground optically active exciton state (A:$1s$), the linewidth at half maximum at a temperature $T\approx 4$~K is a few meV, which is close to the exciton radiative broadening~\cite{PhysRevX.7.021026}. With increasing the temperature (at $T \lesssim 30\ldots 50$~K), the exciton line broadens almost linearly with the slope ratio of $5\times 10^1 \ldots 10^2$ $\mu$eV/K. This qualitatively and quantitatively corresponds to the broadening of the exciton resonance due to interaction with long-wavelength acoustic phonons~\cite{PhysRevLett.119.187402,shree2018exciton}, which confirms insignificant role of a static disorder in high-quality TMDC-based heterostructures. It allows us to analyze the exciton transport within the framework of semiclassical models assuming free propagation of excitons and their relatively rare scattering by phonons. The effects of localization of excitons in such structures can be neglected.\footnote{We leave moire heterostructures outside the scope of this work. In such systems, exciton transport can be qualitatively different.}

In the semiclassical approach, the key parameter that determines the  propagation regime of quasiparticles is the ratio of the exciton lifetime $\tau_r$ and its scattering time $\tau_p$. In typical experiments performed under conditions far from the exciton gas degeneracy
\begin{equation}
\label{tau:relation}
\tau_r \gg \tau_p.
\end{equation}
Therefore, the exciton experiences many collisions during its lifetime~\cite{PhysRevLett.124.166802,Glazov:2022aa,PhysRevLett.127.076801,wagner:trions,PhysRevLett.132.016202,Chernikov:2023aa}. Thus, the propagation of excitons on time scales $t\sim \tau_r$ obeys the diffusion law. The mean square displacement of excitons over time $t$ is described by the relation
\begin{equation}
\label{diffusion}
\Delta \sigma^2(t) = 2Dt, \quad t \gg \tau_p,
\end{equation}
where the diffusion coefficient $D$ can be estimated as
\begin{equation}
\label{D:est}
D = \frac{k_B T}{M}\tau_p,
\end{equation}
and $M$ is the translational mass of the exciton. An order of magnitude estimates give $D\sim 1\ldots 10$~cm$^2/$s. Combined with lifetimes of an exciton ensemble in the range of tens to hundreds of picoseconds, it gives the characteristic diffusion spreading of the exciton cloud $\Delta \sigma(\tau_r)$ of several microns. This situation is typical in many experiments, see~\cite{Chernikov:2023aa} and references therein. In a number of works~\cite{PhysRevB.104.165302,https://doi.org/10.48550/arxiv.2204.09760,fowlergerace2023transport}, however, excitons spread by $\sim 100$~$\mu$m. Also, the above estimates cannot explain the ultrafast propagation of excitons in the first few picoseconds after an excitation, observed, for example, in Ref.~~\cite{PhysRevB.107.045420}: the corresponding diffusion coefficient at the initial stage exceeds $\sim 1000$~cm$^2/$s.

As the pump power and, accordingly, the density of photogenerated excitons increases, exciton-exciton interactions come into play. In both mono- and bilayers of TMDCs, the exciton-exciton annihilation (the Auger effect) plays an important role, where one of the two colliding excitons recombines, transferring its energy and momentum to the second one~\cite{2053-1583-3-3-035011,Moody:16,Manca:2017aa,PhysRevX.8.031073,Lin:2021uu}. It leads to a reduction in $\tau_r$ and a non-exponential decay of the exciton luminescence intensity as a function of time, as well as an effective increase in the observed exciton diffusion coefficient~\cite{PhysRevLett.120.207401,leon2019hot}. The exciton-exciton repulsion is also significant in bilayer structures, which also results in an increase in the diffusion coefficient~\cite{Tagarelli:2023aa,PhysRevLett.132.016202}. According to rough estimates, at not too low temperatures of the exciton gas ($T\gtrsim 10$~K), these effects control the exciton transport up to the Mott transition densities, which is confirmed, for example, by experiments on the reconstructed MoSe$_2$/WSe$_2$ heterobilayer~\cite{PhysRevLett.132.016202}. The joint propagation of excitons and electron-hole plasma~\cite{PhysRevLett.132.016202,zipfel2019exciton}, as well as purely nonequilibrium phenomena associated with the appearance of drift fluxes of excitons due to the Seebeck effect and phonon drag or phonon wind~\cite{Perea-Causin:2019aa,PhysRevB.100.045426} may additionally play a role under such conditions.

According to estimates, at least at low temperatures $T\lesssim 4$~K, there is a fairly wide range of exciton densities $N$, where the exciton gas becomes degenerate while the Auger effect is insignificant. Under the condition~\cite{PhysRevLett.105.070401,PhysRevLett.105.070401,Fogler2014}
\begin{equation}
\label{BKT}
T < T_{BKT}, \quad k_B T_{BKT} \sim \frac{\hbar^2 N}{M},
\end{equation}
the Berezinskii-Kosterlitz-Thouless transition should occur, and the excitonic liquid should become superfluid. In this case, the propagation of excitons becomes nondiffusive, since scattering in a superfluid liquid is suppressed. The spread of excitons is described by a linear law~\cite{Kuznetsov:2020aa}
\begin{equation}
\label{linear}
\Delta\sigma(t) = v t,
\end{equation}
where the speed $v$ is on the order of magnitude the speed of sound (Bogolyubov excitations) $c_s$ in the condensate. For weakly interacting excitons
\begin{equation}
\label{sound:cond}
c_s = \sqrt{\frac{U_0 N}{M}},
\end{equation}
where $U_0>0$ is the exciton-exciton repulsion constant. The product $U_0N$ is the chemical potential of excitons. It is considered small compared to the exciton binding energy and, naturally, compared to the band gap. Therefore, the characteristic velocities $c_s$ are significantly less than the ``Kane'' velocity $v_{cv} = p_{cv}/m_0$ ($p_{cv}$ is the matrix element of the momentum operator, $m_0$ is the mass of the free electron), which corresponds to the interband matrix element of the velocity operator in the crystal.\footnote{In reality, the spread velocity may be even lower and associated with the formation of vortices, cf.~\cite{suris_superfluid}.} The latter is several $10^{-3}$ -- $10^{-2}$ of the speed of light in vacuum. Under experimental conditions~\cite{Aguila:2023aa} the value of $c_s$ does not exceed $10^{-3}c$, but within a few picoseconds the excitons spread over tens of microns, which corresponds to $v \approx 6\times 10^{-2}c$. Therefore, the superfluid propagation of excitons cannot explain the experiment~\cite{Aguila:2023aa}.

\section{Exciton and photon transport}\label{sec:model}

The  ``generally accepted'' transport scenarios outlined above cannot, as noted, explain a number of experimental results on ultrafast exciton propagation. In this section we analyze alternative scenarios due to exciton-light interaction.

To begin with, we note that the interaction of the exciton with the modes of the  electromagnetic field significantly renormalizes its dispersion even for a free monolayer (without hBN and substrate). The renormalization is mainly experienced by the longitudinal exciton, whose microscopic dipole moment (oscillating at the exciton resonance frequency $\omega_{\rm exc} = \mathcal E/\hbar$) is parallel to its wave vector $\mathbf k$. Its dispersion can be written in the form~\cite{PhysRevB.41.7536,goupalov98,ivchenko05a,glazov2014exciton,Yu:2014fk-1,PSSB:PSSB201552211}
\begin{subequations}
\label{long:range}
\begin{equation}
\label{E:long}
\mathcal E_L(\mathbf k) = \mathcal E_L(0)+ \frac{\hbar \Gamma_0}{\mathcal E_L(0)} c k.
\end{equation}
Here $\mathcal E_L(0)$ is the exciton energy at zero wave vector in the plane of the layer, calculated taking into account the direct Coulomb interaction and short-range electron-hole exchange interaction, $\hbar\Gamma_0$ is the radiative damping of the exciton (at $k =0$) expressed in the energy units, and the  $\hbar^2 k^2/2M$ contribution is neglected. Equation~\eqref{E:long} is valid at $k\gg\omega_{\rm exc}/c$ and describes the contribution of the long-range exchange interaction between an electron and a hole to the exciton energy. The presence of the linear dispersion can be interpreted as a result of resonant excitation transfer~\cite{agranovich:galanin}. It follows from Eq.~\eqref{E:long} that the group velocity of the longitudinal exciton
\begin{equation}
\label{v:long}
\mathbf v_L(\mathbf k) = \frac{\mathbf k}{k} \times \frac{\hbar \Gamma_0}{\mathcal E_L(0)} c.
\end{equation}
\end{subequations}
A screening of the long-range exchange interaction in van der Waals heterostructures~\cite{prazdnichnykh2020control} leads to a slight decrease of $\mathcal E_L(\mathbf k)$ and $\mathbf v_L(\mathbf k)$ as compared to Eqs.~\eqref{long:range}. It follows from~\eqref{v:long}  that the speed of longitudinal exciton propagation  is $\sim 10^{-3} c$,\footnote{As for any quasiparticles with linear dispersion, the scattering of a longitudinal exciton on static defects and phonons at $k\to 0$ is suppressed, for example, due to a decrease of the density of states.} which, as noted in~\cite{Aguila:2023aa}, is not enough to explain the experiment.

The ultrafast propagation of excitons discovered in a number of experiments, however, can be associated with the propagation of the electromagnetic field in the waveguide modes, which can be formed in van der Waals heterostructures. It is easy to calculate the dispersion and propagation velocities of waveguide modes for the model system presented in Fig.~\ref{fig:Scheme} within the framework of the general theory of planar waveguides~\cite{Tamir:1988,Ebeling1993}. To that end, we introduce standard notations $n_c = 1$ for the refractive index of air (cap layer), $n_s = n_{\rm SiO_2} = 1.46$ for the refractive index of the SiO$_2$ substrate and $n_f = n_{\rm hBN}= 2.2$ for the refractive index of a hexagonal boron nitride film (we neglect the anisotropy of its optical properties as well as the frequency dispersion of the refractive indices for simplicity)~\cite{PhysRevMaterials.2.011001}. We also neglect the contribution to the reflection of light from the TMDC monolayer.\footnote{It can be taken into account as a small renormalization of the thickness of the boron nitride layer $d$.} Since the relation is satisfied
\[
n_f>n_s>n_c,
\]
a waveguide mode can indeed form in the boron nitride layer. The waveguide asymmetry parameters for TE and TM modes are written in the form
\begin{equation}
\label{asymmetry}
\alpha_{\rm TE} = \frac{n_s^2 - n_c^2}{n_f^2 - n_s^2}, \quad \alpha_{\rm TM} = \alpha_{\rm TE} \frac{n_f ^4}{n_c^4}.
\end{equation}
The dispersion law for TE modes can be represented as~\cite{Tamir:1988,Ebeling1993}
\begin{multline}
\label{disper:TE}
V\sqrt{1-b} = \pi m + \arctan{\sqrt{b/(1-b)}}\\
+\arctan{\sqrt{(b+\alpha_{\rm TE})/(1-b)}},
\end{multline}
where
\begin{equation}
\label{notations}
V = \frac{\omega}{c}d \sqrt{n_f^2 - n_s^2}, \quad b = \frac{(ck/\omega)^2 - n_s^2}{n_f^2 - n_s ^2}.
\end{equation}
Here $\omega$ is the mode frequency, $k$ is the absolute value of the wave vector in the layer plane, and $m=0,1,2,\ldots$ enumerates the modes.

\begin{figure}[t]
 \includegraphics[width=\linewidth]{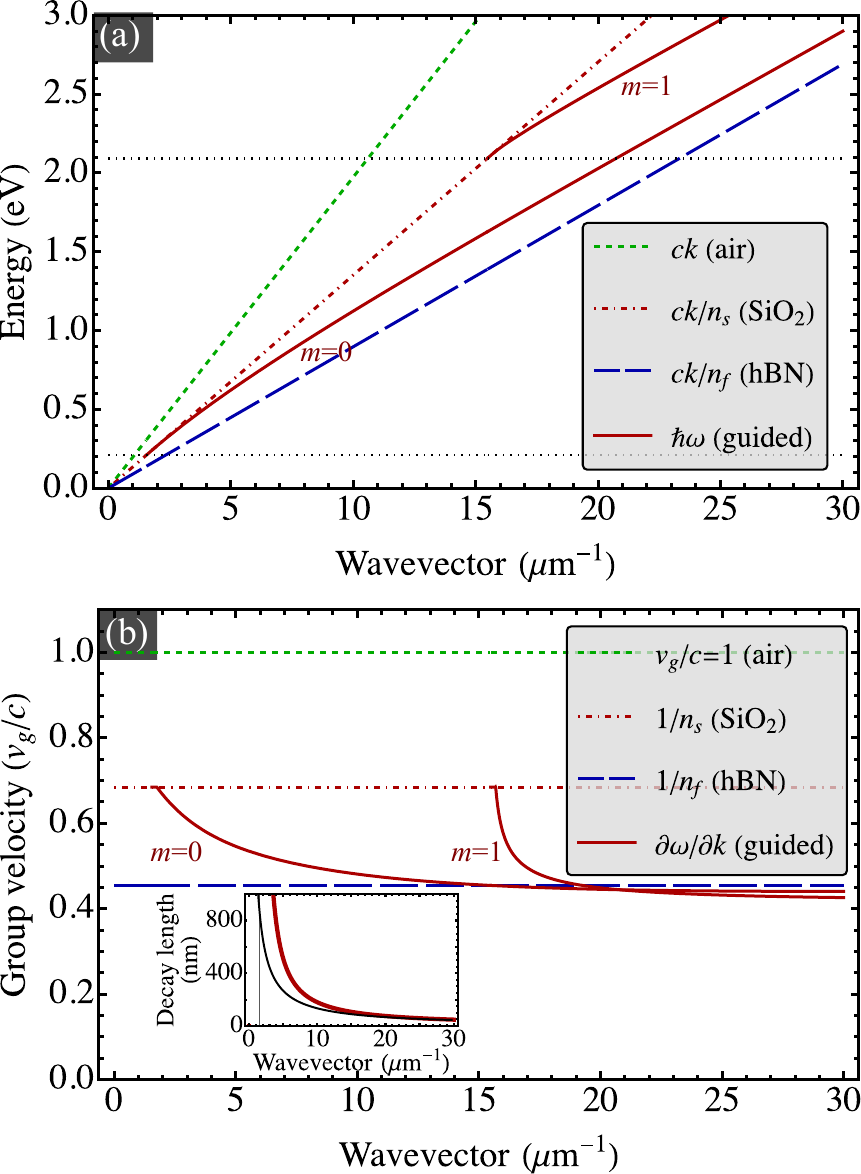}
  \caption{(a) Dispersion of two lowest in energy TE-waveguide modes (solid curves) calculated by solving Eq.~\eqref{disper:TE} for the structure depicted in Fig.~\ref{fig:Scheme}, as well as the light dispersion in the air (dotted curve), the waveguide hBN layer (dashed curve) and in the SiO$_2$ substrate (dash-and-dot curve). Horisontal dotted lines show the cut-off energies of the modes, Eq.~\eqref{cutoff} and Fig.~\ref{fig:CutOff}. (b) Group velocity of the modes calculated after Eq.~\eqref{vg}. Inset shows the decay length of the waveguide mode into SiO$_2$ layer: exact calculation according to Eq.~\eqref{penetration} is shown by the solid red line and the large wavevector approximation by the thin black line. hBN layer thickness $d=200$~nm. Refractive indices: $n_{\rm hBN} = 2.2$, $n_{\rm SiO_2}=1.46$.
}\label{fig:Modes}
\end{figure}

\begin{figure}[b]
 \includegraphics[width=\linewidth]{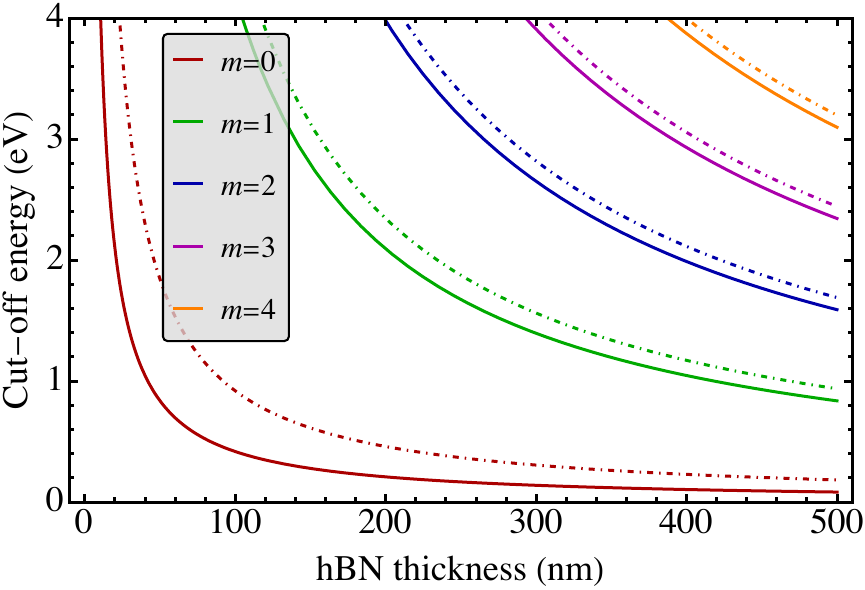}
  \caption{Cut-off energies of the five lowest in energy modes in the structure depicted in Fig.~\ref{fig:Scheme} as function of the hBN layer thickness calculated after Eq.~\eqref{cutoff}. TE-modes are shown by the solid curves and TM modes by the dash-and-dot curves. Refractive indices: $n_{\rm hBN} = 2.2$, $n_{\rm SiO_2}=1.46$.
}\label{fig:CutOff}
\end{figure}

Figure~\ref{fig:Modes}(a) shows the dispersion of the two lowest energy TE-polarized waveguide modes in the structure under consideration (solid curves marked $m=0$ and $m=1$), as well as dispersion lines for light in the air (dotted line), in hBN (dashed line), and in SiO$_2$ (dash-and-dot line). The dispersion is calculated by Eq.~\eqref{disper:TE}. It can be seen that the waveguide modes are present in a wide range of wavevectors and energies. Their energy varies from the photon energy in the substrate layer ($\hbar c k/n_s$) at small wavevectors, where the field is mainly concentrated in the SiO$_2$ layer, to the energy $\hbar c k/n_f$ at large $k$, where the field is localized in the hBN layer. The mode cut-off energies at small $k$ are determined by the condition of ``detachment'' of the dispersion curve of the waveguide mode from the light cone of the substrate, which corresponds to $b=0$ in~\eqref{notations}. We obtain from Eq.~\eqref{disper:TE} 
\begin{equation}
\label{cutoff}
\frac{\omega^{\rm cut-off}_{\rm TE,TM}}{c} = \frac{\sqrt{n_f^2 - n_s^2}}{d }\left(\pi m+ \arctan{\sqrt{\alpha_{\rm TE,TM}}}\right).
\end{equation}
For the selected structure parameters, the cut-off energies of the two lowest TE-modes are represented in Fig.~\ref{fig:Modes}(a) by horizontal dotted lines. The dependence of the cut-off energies on the thickness of the boron nitride layer is illustrated in Fig.~\ref{fig:CutOff} (solid curves correspond to the TE-modes, dash-and-dot curves correspond to the TM-modes). The field penetration depth into the SiO$_2$ substrate calculated according to
\begin{equation}
\label{penetration}
\ell = \left[ k^2 - \left(n_s\frac{\omega}{ c}\right)^2\right]^{-1/2},
\end{equation}
is shown in the inset to Fig.~\ref{fig:Modes}(b) by a solid dark red curve. The thin black curve is the asymptotic~\eqref{penetration} for large $k$, where $\omega = ck/n_f$.

Figure~\ref{fig:Modes}(b) shows the group velocity of propagation of the two lowest energy waveguide TE-modes ($m=0,1$)
\begin{equation}
\label{vg}
\mathbf v_g = \frac{\partial \omega}{\partial \mathbf k},
\end{equation}
calculated from the dispersion found from Eq.~\eqref{disper:TE}. The group velocity of waveguide modes varies in the range from the speed of light in the SiO$_2$ substrate (at small wave vectors) to approximately the speed of light in hexagonal boron nitride (at large values of $k$). It is several tenths of the speed of light in a vacuum. Thus, the structure shown in Fig.~\ref{fig:Scheme} supports ultrafast (compared to excitons) optical waveguide modes.

As one can see from Fig.~\ref{fig:Modes}, these modes are located below the light cone in a vacuum and cannot be directly excited by the light incident on the sample. Under experimental conditions, the waveguide modes can be excited by the light scattering on inhomogeneities of the structure (the characteristic dimensions of the samples under study are 10 \ldots 100~$\mu$m) or via the secondary radiation of excitons excited by the incident light. The scattering from roughnesses can be suppressed near the cut-off frequency due to the large depth of field penetration into SiO$_2$ [see inset in Fig.~\ref{fig:Modes}(b)]. Nonlinear and threshold effects observed experimentally at the ultrafast exciton propagation can be associated with changes in the light scattering conditions (for example, due to a light-induced shift of exciton dispersion) or with an increase of the number of photons entering the waveguide mode. Light, propagating by waveguide modes, can generate excitons in a monolayer far from the excitation spot, which, in turn, generate luminescence, detected experimentally at the exciton resonance frequency. So the process
\begin{multline}
\label{process}
\mbox{incident photon} \\
\rightarrow \mbox{scattering (roughnesses, excitons)} \\
\rightarrow \mbox{photon in a waveguide mode}\\
\rightarrow \mbox{exciton/scattering}
\rightarrow \mbox{secondary photon}
\end{multline}
may be responsible for the apparent ultrafast propagation of excitons. In fact, the energy is transferred by light in the waveguide mode, and the propagation delay consists of the actual delay in light propagation and the delays related to the formation of the secondary luminescence. We emphasize that if the waveguide mode is populated due to the emission of photoexcited excitons (light scattering by excitons), then a secondary photon at the exciton frequency can be formed when the waveguide photon is scattered on inhomogeneities of the structure.

A diffusive propagation of excitons and ultrafast transfer of photons by the waveguide modes leads to different spatial profiles of the secondary luminescence intensity distribution. For simplicity, let us consider a one-dimensional case. For the diffusion, the distribution has a Gaussian shape:
\begin{subequations}
\label{profile}
\begin{equation}
\label{profile:Diff}
I_{\rm diff}(x, t) \propto \frac{1}{\sqrt{4\pi D t}}\exp{\left(-\frac{x^2}{4Dt} \right)} .
\end{equation}
On the contrary, for the proposed process~\eqref{process} the light pulse propagates with a group velocity $v_g$~\eqref{vg}, and the secondary luminescence appears with a delay for a relaxation time wherever the waveguide mode has already ``passed'':
\begin{equation}
\label{profile:wg}
I_{\rm wg}(x,t) \propto \Theta(|x|-v_gt).
\end{equation}
Here $\Theta(x)$ is the Heaviside step function.
The last expression holds true at times significantly exceeding the delay time. Scattering and absorption of the waveguide mode leads to the exponential $\exp{(-|x|/L)}$ in the intensity distribution, where $L$ is the attenuation length of the waveguide mode.
\end{subequations}

\section{Conclusion}

In this paper we have presented a brief review of the exciton propagation regimes in semiconductor heterostructures based on atomically thin semiconductors -- transition metal dichalcogenides. The features of diffusive and superfluid propagation of quasiparticles have been presented. On the basis of the calculation of photon modes in van der Waals heterostructures, a mechanism of ultrafast propagation has been proposed. It is related to the formation of waveguide modes in a layer of hexagonal boron nitride, their excitation by the light scattering on structure inhomogeneities or excitons, their propagation, subsequent generation of excitons and their emission. The mechanism can lead to an ultrafast -- with speeds constituting tenths of the speed of light in a vacuum -- energy transfer. This effect can lead to the apparent ultrafast propagation of excitons.

The authors do not claim that the proposed mechanism is fully responsible for all observations of ultrafast exciton transport in van der Waals heterostructures. In particular, we do not pretend for the full description of the experiments~\cite{PhysRevB.104.165302,https://doi.org/10.48550/arxiv.2204.09760,fowlergerace2023transport,PhysRevB.107.045420,Aguila:2023aa}. Noteworthy, that in hBN-less structres described waveguide modes do not form. In structures with a more complex hBN/SiO$_2$/Si substrate, the refractive index of silicon exceeds that of the remaining layers. Hence, the waveguide mode will be attenuated by tunneling into the silicon substrate, but this effect is suppressed at relatively large wavevectors. On the other hand, if the damping of the waveguide mode is small, and for a certain wavevector, the resonance condition of the exciton energy and the waveguide mode is satisfied, then a strong coupling regime is also possible. It is accompanied by the formation of exciton-polaritons similar to those considered in~\cite{Guo:2022aa,PhysRevB.98.161113} for structured substrates. In this regime, significant renormalization of the dispersion and, naturally, changes in the regimes of propagation of quasiparticles are expected~\cite{PhysRevB.108.104202}.

We note in conclusion that a possibility of waveguide mode formation should be taken into account for the analysis of exciton transport in van der Waals heterostructures. Further experiments and detailed modeling can finally clarify the origin of ultrafast propagation of excitons or photons in these interesting objects.

Authors thank  Andr\'es Granados del \'Aguila for valuable discussions. This work was supported by RSF grant No. 23-12-00142.

%

\end{document}